\documentclass[preprint,aps,tightenlines,floatfix]{revtex4}

\usepackage{graphics}
\usepackage{bm}
\usepackage{amsmath}
\usepackage{amssymb}

\begin{document}

\title{A simple functional form for proton-nucleus total reaction cross sections}

\author{K. Amos}
\email{amos@physics.unimelb.edu.au}

\author{P. K. Deb}
\email{deb@physics.unimelb.edu.au}

\affiliation{School of Physics, The University of Melbourne, Victoria,
3010, Australia}

\date{\today}

\begin{abstract}
A simple functional form has been found that gives a good representation
of the total reaction cross sections for the scattering of protons
from (15) nuclei spanning the mass range ${}^{9}$Be to ${}^{238}$U 
and for proton energies ranging from 20 to 300 MeV.
\end{abstract}

\maketitle

\section{introduction}
Total reaction cross sections of nucleon scattering from nuclei (stable
as well as radioactive) are the most basic nuclear observables~\cite{Di80}
being the sum of all the reaction processes induced during nuclear collisions.
Reliable methods of predicting these values accurately and quickly  are needed for 
a variety of applications~\cite{Ra98}; applications which include the production of 
radioactive ion beams~\cite{Ge95}, studies of the relative abundances of the nuclei in 
galactic cosmic rays~\cite{We87}, as well as for radiation shielding design for future 
space exploration vehicles~\cite{Wi82}.  Some others of quite current interest concern the 
transmutation of long lived radioactive waste into shorter lived products using accelerator 
driven systems (ADS)~\cite{Ik02,Gu02,Sh02} and in predicting dosimetry for patients
in radiation therapy~\cite{Ch00}.
To be able to specify those total reaction cross sections in a simple functional form 
then has great utility for any associated evaluation such as of nucleon production in
spallation calculations.  

Recently~\cite{De01} nucleon-nucleus total reaction 
cross sections to 300 MeV for $^{12}$C and $^{208}$Pb have been predicted in good 
agreement with data.  Cross sections were evaluated using $g$-folding optical 
potentials formed by full folding realistic two nucleon ($NN$) interactions with 
credible nucleon-based structure models of those nuclei.  
In particular, the ground state of ${}^{12}$C was specified from  a 
complete $(0+2)\hbar\omega$ shell model evaluation~\cite{De01}, while that of 
${}^{208}$Pb was obtained from a Skyrme-Hartree-Fock calculation~\cite{Br00}.
The effective interactions 
at each energy in the range to 300 MeV were defined from the $NN$ $g$ matrices (solutions 
of the Bruckner-Bethe-Goldstone equations) of the free $NN$ (BonnB) interactions.
Consequently they vary with energy and the medium density.  
The resulting $nA$ optical potentials are complex, energy dependent and non-local.
All details of the approach and numerous demonstrations of its successful use with 
targets spanning the entire mass range (3 to 238) are given in the recent 
review~\cite{Am00}.  But each such prediction of nucleon scattering is an involved 
calculation that culminates with
use of large scale computer programs, and those of DWBA98~\cite{Ra98a,Am00} in recent
studies.

It would be very utilitarian if aspects of such scattering were indeed well approximated by 
a simple convenient functional form.  We demonstrate herein  that for the total reaction
cross sections such a form may exist for nuclei spanning the mass range.  In all we report 
results from 15 nuclei ranging in mass from 9 to 238.  Also we  have studied proton scattering 
for energies ranging from 20 MeV to 300 MeV.

%%%%%%%%%%%%%%%%%%%%%%%%%%%%%%%%%%%%%%%%%%%%%%%%%%%%%%%%%%%%%%%%%%%%%%%%%%%%%%%%%%%%%%

\section{A simple functional form for ${\small p}-A$ total reaction cross sections}

Initially we have used
a set of $g$-folding optical potential calculations to  specify
proton-nucleus  scattering $S$ matrices, i.e with energies $E\propto k^2$, 
\begin{equation}
S^{\pm}_l \equiv S^{\pm}_l(k) = e^{2i\delta^{\pm}_l(k)} =
\eta^{\pm}_l(k)e^{2i\Re\left[ \delta^{\pm}_l(k) \right] }\ ,
\end{equation}
where $\delta^\pm_l(k)$ are the (complex) scattering phase shifts and $\eta^{\pm}_l(k)$ 
are the moduli of the $S$ matrices. The superscript designates $j = l\pm 1/2$.
Therefrom total reaction cross sections are predicted by
\begin{eqnarray}
\sigma_R(E)  &=&  \frac{\pi}{k^2} \sum^{\infty}_{l=0} \left\{ (l+1)
\left[ 1 - \left( \eta^+_l \right)^2 \right] + l \left[ 1 - \left(
\eta^-_l \right)^2 \right] \right\}\nonumber\\
&=& \frac{\pi}{k^2}\sum^{\infty}_{l=0} \sigma_l^{(R)}(E)\ .
\label{xxxx}
\end{eqnarray}
The partial reaction cross sections $\sigma_l^{(R)}(E)$ so found we now take as 'data',
and such will be so specified hereafter. We note that there are discrepancies between
the predictions of total reaction cross sections found using the $g$-folding approach
and actual data due to limitations in that approach, e.g. in the structure model used to 
describe the ground state densities of some nuclei. However there have been so many 
successes with the approach~\cite{De01,Am00,Ka02} that we believe the functional form 
developed here on the basis of $g$-folding potential calculated values is pertinent
with but minor refinement (of the values of the three parameter values) to reproduce
actual measured total reaction cross sections.

As will be evident from the figures presented
later, the results clearly suggest that such partial reaction cross sections 
can be described by a simple functional form such as
%%%%%%%%%%%%%%%%%%%%%%%%%%%%%%%%%%%%%%%%%%%%%%%%%%%%%%%%%%%%%%%%
\begin{equation}
\sigma_l^{(R)}(E) = (2l+1) \left[1 + e^{\frac{(l-l_0)}{a}}\right]^{-1}
 + \epsilon (2l_0 + 1)
 e^{\frac{(l-l_0)}{a}}
\left[ 1 + e^{\frac{(l-l_0)}{a}} \right]^{-2}\ ,
\label{Fnform}
\end{equation}
%%%%%%%%%%%%%%%%%%%%%%%%%%%%%%%%%%%%%%%%%%%%%%%%%%%%%%%%%%%%%%%%
in which the parameters,  $l_0(E,A)$, $a(E,A)$, and $\epsilon(E,A)$, should vary smoothly 
with energy and mass.

The summation giving the total reaction cross section can be limited to a value $l_{max}$ 
and the associated form tends appropriately to the known high energy limit.  With increasing 
energy, $l_{max}$ becomes so large that the exponential fall  of the functional form, 
Eq.~\ref{Fnform}, can be approximated as a straight vertical line ($l_0 = l_{max}$).
In that case, the total reaction cross section equates to the area of a triangle and
\begin{equation}
\sigma_R \Rightarrow \frac{\pi}{2 k^2} l_{max}(2l_{max} + 1)
 \approx \frac{\pi}{k^2} l_{max}^2\ .
\end{equation}
Then with $l_{max} \sim kR$, at high energies, $\sigma_R \Rightarrow \pi R^2$;
the geometric cross section as required.

%%%%%%%%%%%%%%%%%%%%%%%%%%%%%%%%%%%%%%%%%%%%%%%%%%%%%

For any dynamical situation in which numerous partial reaction cross sections contribute 
non-negligibly to the sum in Eq.~\ref{xxxx}, one can use the limit form
\begin{equation}
\sigma_R(E)
= \frac{\pi}{k^2}\sum^{\infty}_{l=0} \sigma_l^{(R)}(E)
\Longrightarrow \frac{\pi}{k^2} \int_0^\infty \sigma^{(R)}(E; \lambda)
\ d\lambda\ .
\label{eqn-integ}
\end{equation}
Then with $x = \frac{\lambda - l_0}{a}$ and using the basic integrals
\begin{eqnarray}
\int_{-l_0/a}^\infty \frac{1}{1 + e^x} dx
&=&
\left[ \int_0^{l_0/a} \frac{1}{1 + e^{-x}} dx
+ \int_0^\infty \frac{e^{-x}}{1 + e^{-x}} dx\right]
\nonumber\\
&=&
 \frac{l_0}{a} + \sum_{n=1}^\infty (-)^{n+1} e^{-\frac{l_0}{a}n}
\nonumber\\
\int_{-l_0/a}^\infty \frac{x}{1 + e^x} dx
&=& \left[ -\int_0^{l_0/a} \frac{x}{1 + e^{-x}} dx
+ \int_0^\infty \frac{xe^{-x}}{1 + e^{-x}} dx\right]
\nonumber\\
&=&
-\frac{1}{2} \frac{l_0^2}{a^2}
+ 1.645 + \sum_{n=1}^\infty
(-)^n \left( 1 + \frac{1}{n^2}\right) e^{-\frac{l_0}{a}}
\ ,
\end{eqnarray}
the integral observable is
\begin{eqnarray}
\sigma^{(R)}(E) &=& \frac{\pi}{k^2} \sum_l \sigma_l^{(R)}(E)
\to \frac{\pi}{k^2} \int_0^\infty \sigma^{(R)}(E;\lambda) \
d\lambda
\nonumber\\
&=& \frac{\pi}{k^2} \Biggl\{
l_0(l_0+1) + 3.29 a^2 + a\epsilon (2l_0+1)
\nonumber\\
&&\hspace*{0.1cm}
+ \sum_{n=0}^\infty e^{-n\frac{l_0}{a}}
(-)^n \left[
2a^2 \left\{ 1+ \frac{1}{n^2}\right\} +
(2l_0+1)\left\{ a\epsilon - \frac{1}{n} \right\}
\right] \Biggr\}
\nonumber\\
&&{\mathop \to_{n=1}}\
\frac{\pi}{k^2}
\Biggl\{
l_0(l_0+1) + 3.29 a^2 + a\epsilon (2l_0+1)
\nonumber\\
&&\hspace*{3.0cm}
+  \left[(2l_0 + 1) \left\{1 - a\epsilon\right\}
- 4a^2  \right] e^{-\frac{l_0}{a}}
\Biggr\}
\label{sig_fn_tot}
\end{eqnarray}
%%%%%%%%%%%%%%%%%%%%%%%%%%%%%%%%%%%%

We have used this (n=1) integral form for all energies and masses and, as could be 
expected, results are not good when the number of significant partial wave terms
is small, as is the case for low energies. But from those comparisons, limit conditions for 
application of the integral representation are evident.

\section{Results}

\begin{figure}
\scalebox{0.7}{\includegraphics{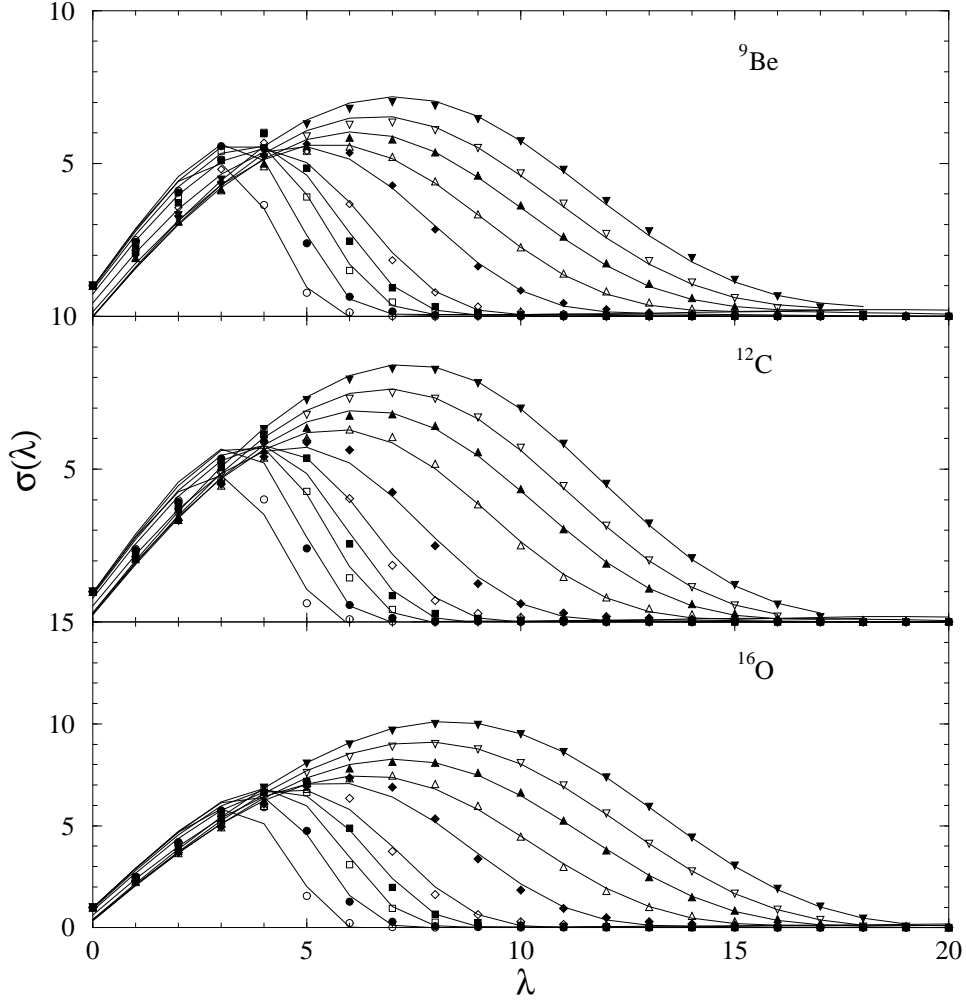}}
\caption{\label{fig-be+c+o}
The total partial reaction cross sections for
proton scattering from ${}^9$Be (top), ${}^{12}$C (middle) and
${}^{16}$O (bottom).}
\end{figure}

\begin{figure}
\scalebox{0.7}{\includegraphics{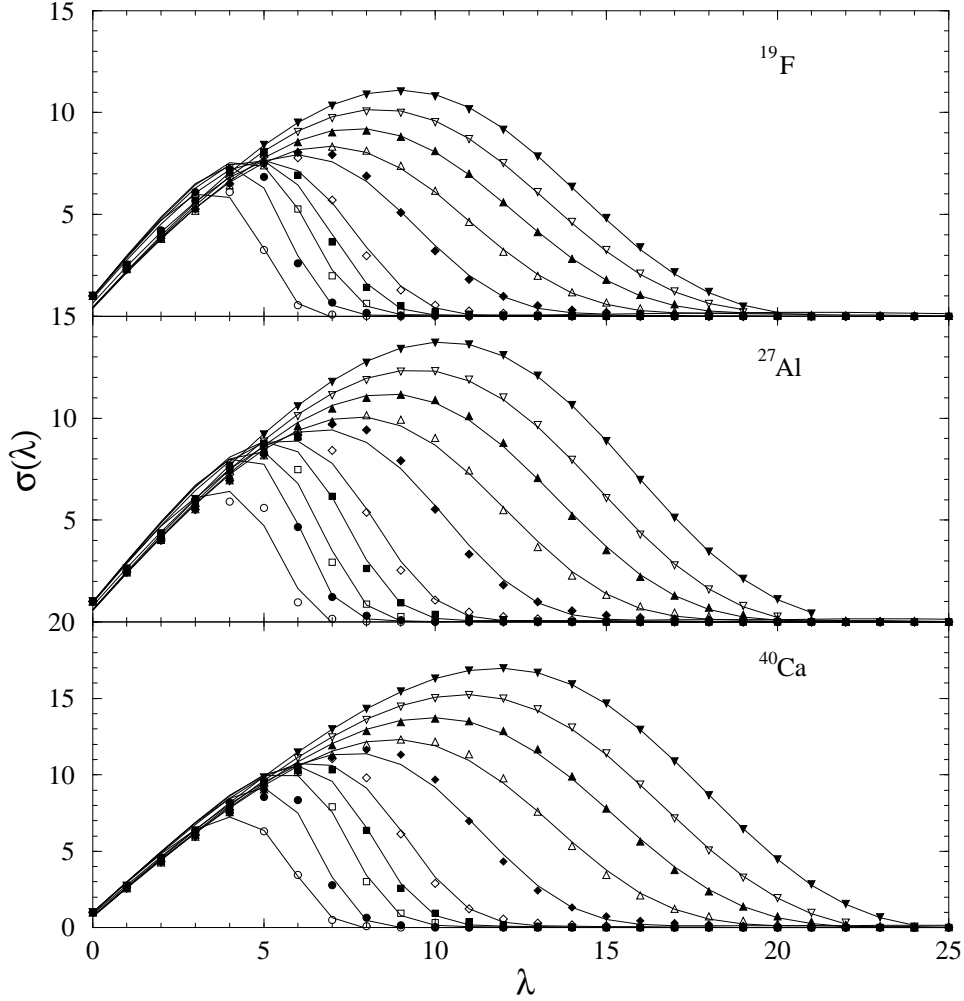}}
\caption{\label{fig-f+al+ca}
Same as Fig.~\ref{fig-be+c+o}, but for
 ${}^{19}$F (top), ${}^{27}$Al (middle) and
${}^{40}$Ca (bottom).}
\end{figure}

\begin{figure}
\scalebox{0.7}{\includegraphics{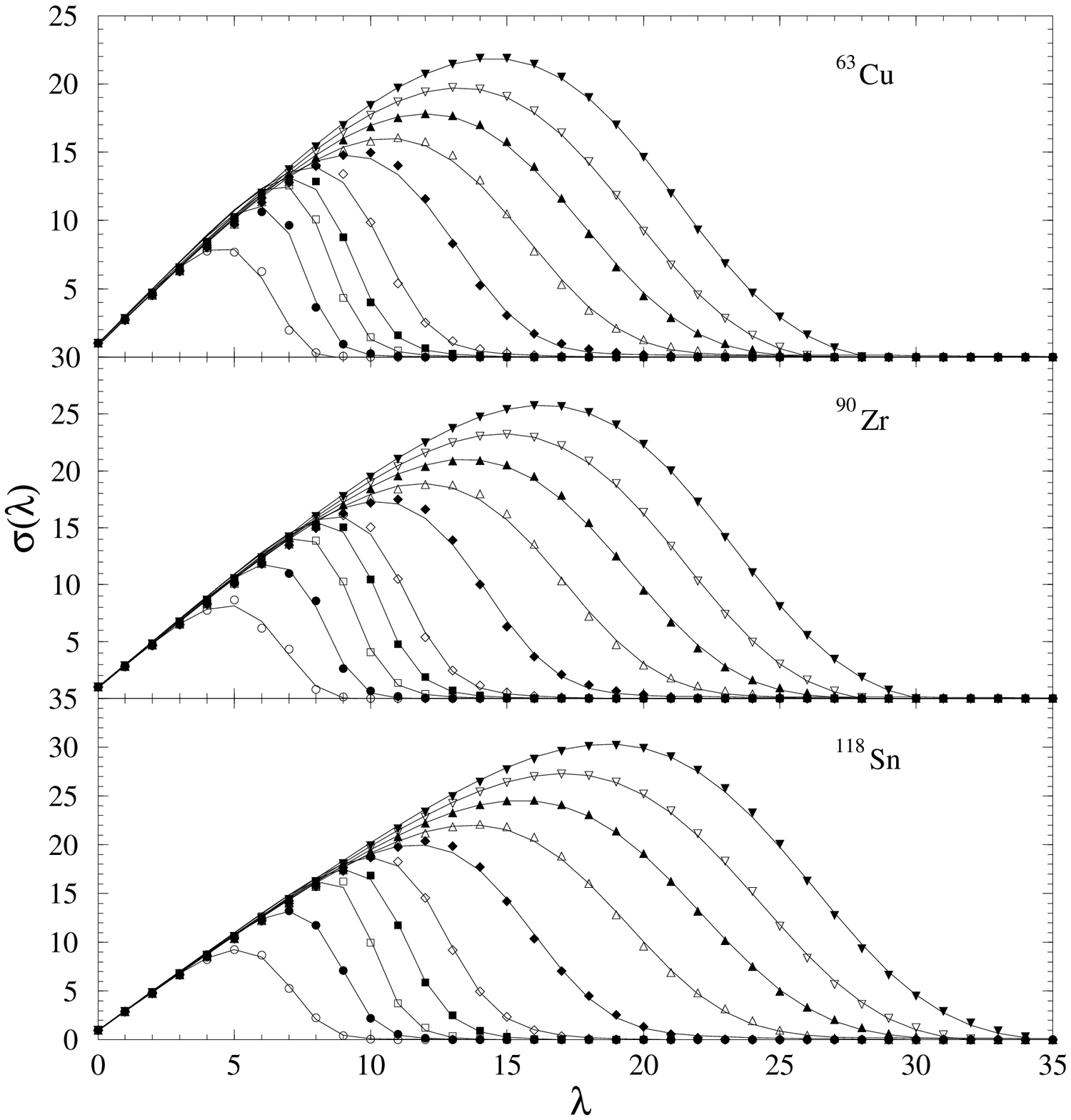}}
\caption{\label{fig-cu+zr+sn}
Same as Fig.~\ref{fig-be+c+o}, but for
 ${}^{63}$Cu (top), ${}^{90}$Zr (middle) and ${}^{118}$Sn (bottom).}
\end{figure}

\begin{figure}
\scalebox{0.7}{\includegraphics{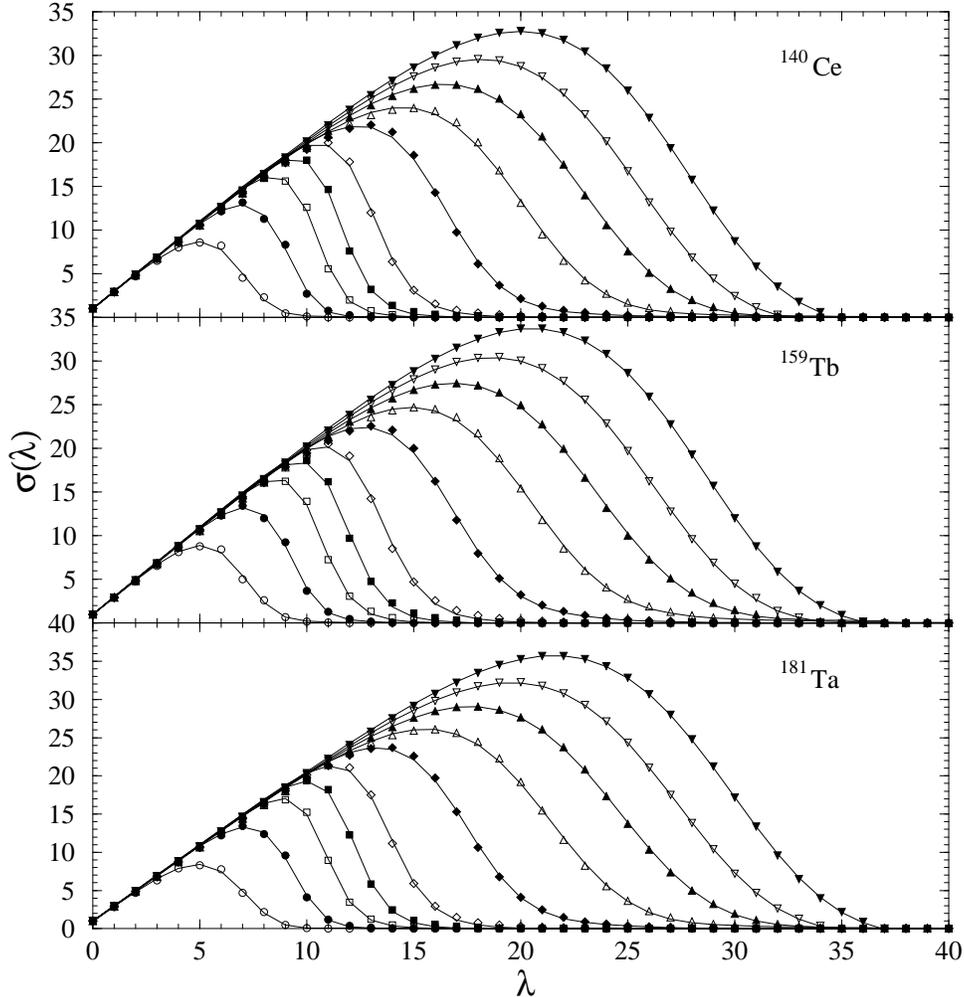}}
\caption{\label{fig-ce+tb+ta}
Same as Fig.~\ref{fig-be+c+o}, but for
 ${}^{140}$Ce (top), ${}^{159}$Tb (middle) and
${}^{181}$Ta (bottom).}
\end{figure}

\begin{figure}
\scalebox{0.7}{\includegraphics{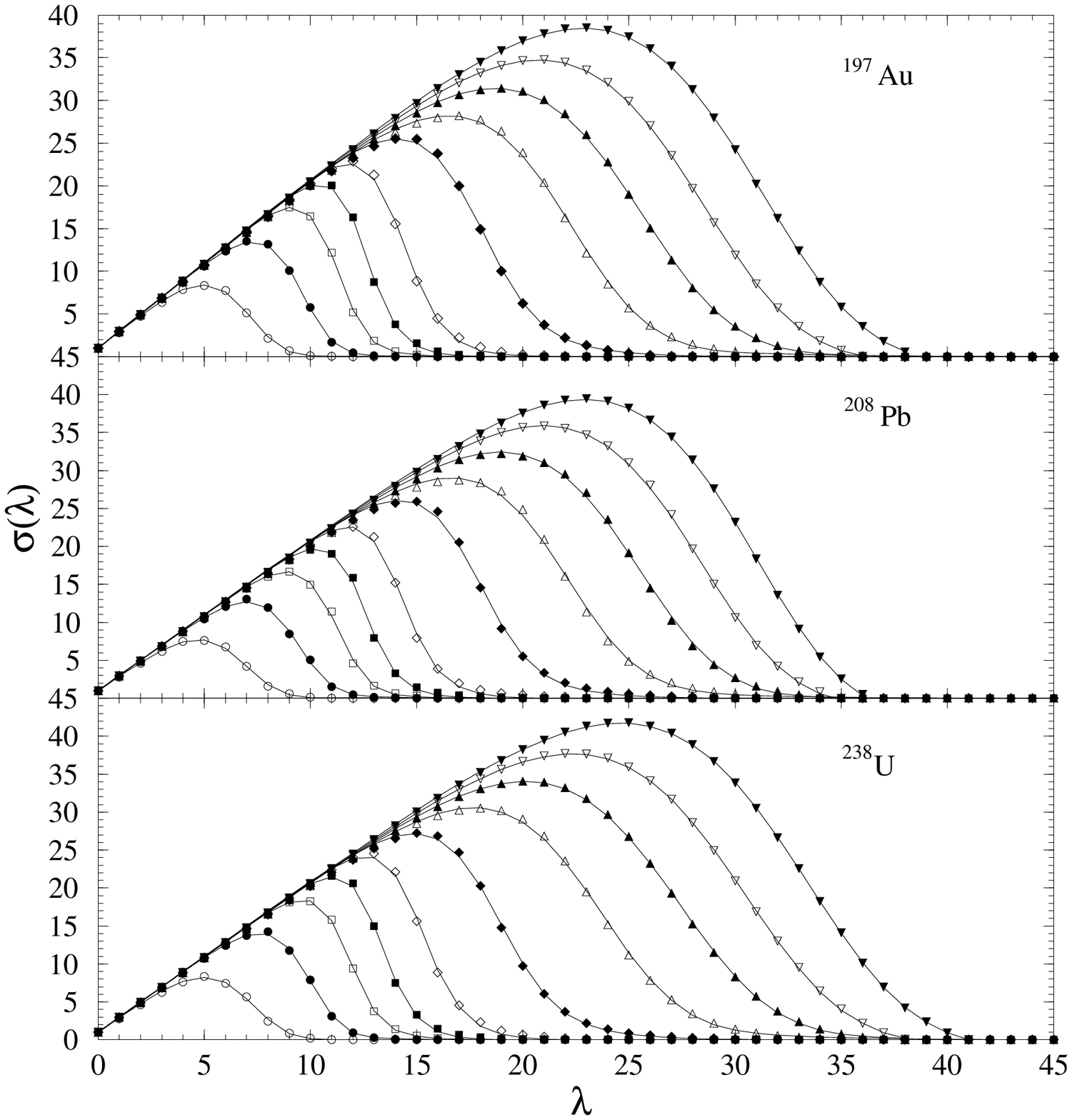}}
\caption{\label{fig-au+pb+u}
Same as Fig.~\ref{fig-be+c+o}, but for
${}^{197}$Au (top), ${}^{208}$Pb (middle) and
${}^{238}$U (bottom).}
\end{figure}

We have calculated the partial reaction cross sections, $\sigma_l$(E),  of the
scattering of protons from 15 different nuclei, namely ${}^9$Be, ${}^{12}$C, ${}^{16}$O, 
${}^{19}$F, ${}^{27}$Al, ${}^{40}$Ca, ${}^{63}$Cu, ${}^{90}$Zr, ${}^{118}$Sn, ${}^{140}$Ce, 
${}^{159}$Tb, ${}^{181}$Ta, ${}^{197}$Au, ${}^{208}$Pb and ${}^{238}$U. The results from
the calculations using the $g$-folding optical potentials (the data) were then 
fit using  the simple functional form given in  Eq.~\ref{Fnform}. The comparisons are displayed in 
Figs.~\ref{fig-be+c+o} through ~\ref{fig-au+pb+u}.  In these figures, the solid lines represent 
the  results calculated using the simple functional form. The dots are the data.
We have considered protons with  
10 different energies for each nucleus and the partial reaction cross sections for
each nucleus are shown in the figures by different symbols according to the proton energy.
Specifically we use the symbol shown after each number for energies of 20 [empty circles], 
30 [filled circles], 40 [empty squares], 50 [filled squares], 65 [empty diamonds], 
100 [filled diamonds], 150 [empty up triangles], 200 [filled up triangles], 250 
[empty down triangles] and 300 [filled down triangles] MeV.  For each mass there is a clear 
progression in the data with energy; one that is similar over the entire mass range
studied and, in fact, was the motivation for our search for an analytic function to 
describe all of that data. 

At low energies, 20 MeV especially and for the light masses most evidently, there are few 
significant  partial wave contributions to the reaction cross sections. The peak of these 
partial wave reaction cross section data thus is not large and occurs at small values of $l$.
Concomitantly, the 3 parameter fit function will involve small values of $l_0$  and the fit 
to the (limited) data set may not be  as uniquely determined as at higher energies,  
particularly as the  associated 
variations of $\sigma_l$ (data and calculated result) are not dominated by the straight line 
factor ($2l+1$) at low $l$-values.  Also the individual data values, for 20 MeV and light mass 
targets, are  more sensitive to details of the actual optical potentials used than 
is the case for the
corresponding  problem of matching the $\sigma_l$ values at higher energies (and target mass) 
with the  3 parameter functional form.  

In Fig.~\ref{fig-be+c+o}, the partial reaction cross sections for proton scattering from 
${}^9$Be (top), ${}^{12}$C (middle) and ${}^{16}$O (bottom) obtained from the functional form 
calculations are compared with our data.  Despite the foregoing comment, the calculated values 
are in excellent agreement  with most of this data. It is clear that the shapes of  
these cross sections for proton scattering from light nuclei and for energies 65 MeV and 
below, differ somewhat from those at 100 MeV and above. Such differences persist with heavier 
targets. Also as the  nuclear mass increases, the linear factor is more evident.
%%%%%%%%%%%%%%%%%%%%%%%%%%% fig1 %%%%%%%%%%%%%%%%%%%%%
%%%%%%%%%%%%%%%%%%%%%%%%%
The partial reaction cross sections for proton scattering from ${}^{19}$F, ${}^{27}$Al and 
${}^{40}$Ca  obtained from the calculations using the  simple functional form are compared
with data in Fig.~\ref{fig-f+al+ca} in the top, middle and bottom panels respectively. 
That data  are quite well reproduced.  But as noted before, only a small number 
of values are of significance at 20 MeV.
%%%%%%%%%%%%%%%%%%%%%%%%%%% fig2 %%%%%%%%%%%%%%%%%%%%%
%%%%%%%%%%%%%%%%%%%%%%%%%
The predictions for the partial reaction cross sections for proton scattering from 
${}^{63}$Cu  to higher mass nuclei are in excellent agreement with our data and such is
evident in Figs.~\ref{fig-cu+zr+sn}, ~\ref{fig-ce+tb+ta} and ~\ref{fig-au+pb+u}.
For these targets,  the dominance of the linear term ($2l+1$) of Eq.~\ref{Fnform} in the fit to 
the low value partial wave data is evident for all energies including  20 MeV since now, even 
at that energy, a reasonable number of partial wave terms contribute significantly to the reaction 
cross sections.  Then  $l_0$ is large enough so that the Wood-Saxon function in Eq.~\ref{Fnform} 
essentially is constant for $0 \le l \le 5$ or more.
%%%%%%%%%%%%%%%%%%%%%%%%%% fig3 %%%%%%%%%%%%%%%%%%%
%%%%%%%%%%%%%%%%%%%%%%%%% fig 4 %%%%%%%%%%%%%%%%%
%%%%%%%%%%%%%%%%%%%%%%% fig 5 %%%%%%%%%%%%%%%%%%%%%%
%%%%%%%%%%%%%%%%%%%%%%%%%

\begin{figure}
\scalebox{0.7}{\includegraphics{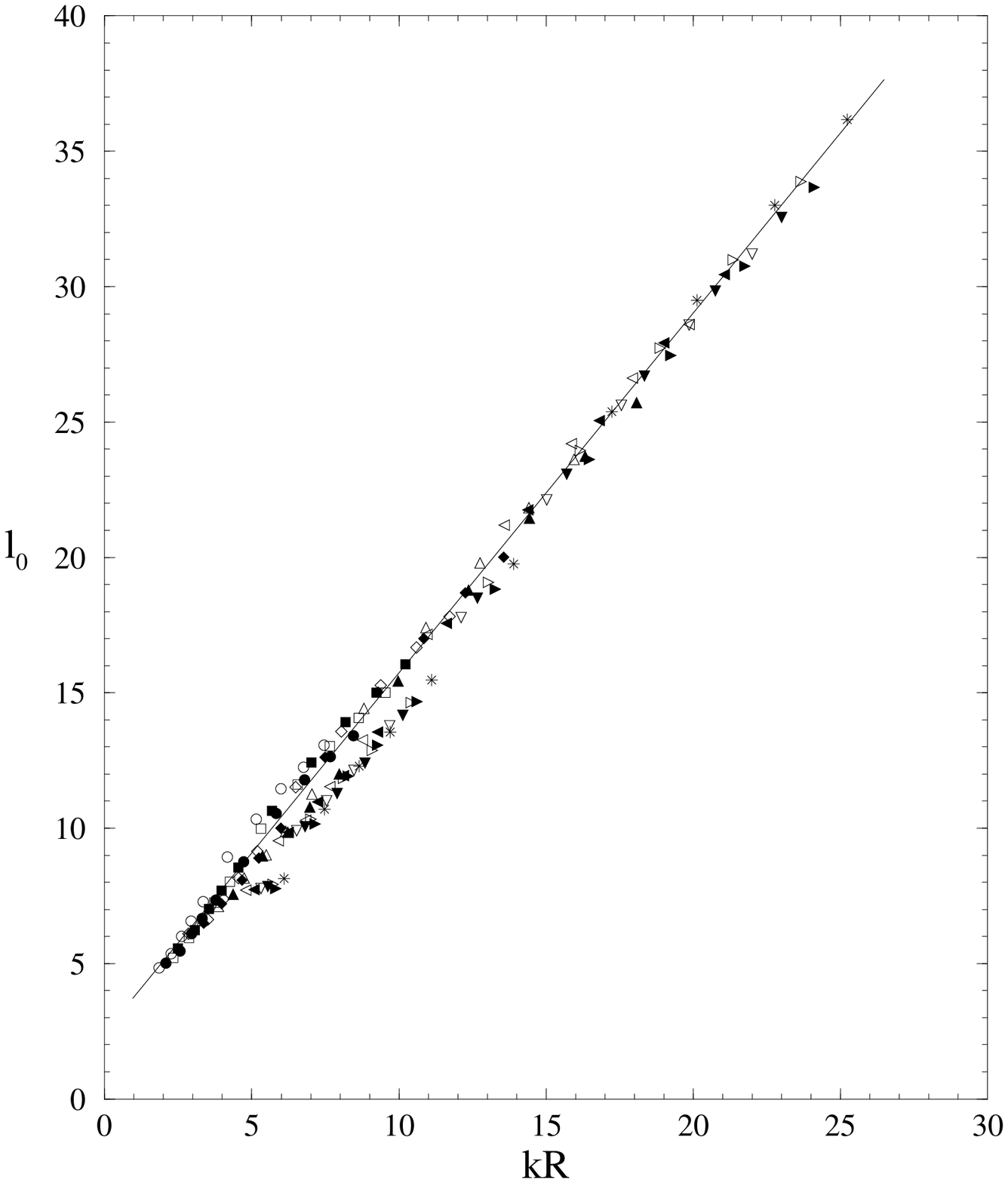}}
\caption{\label{fig-l0}
The parameter $l_0$ of the simple functional form for the
proton-nucleus partial reaction cross sections.}
\end{figure}

\begin{figure}
\scalebox{0.7}{\includegraphics{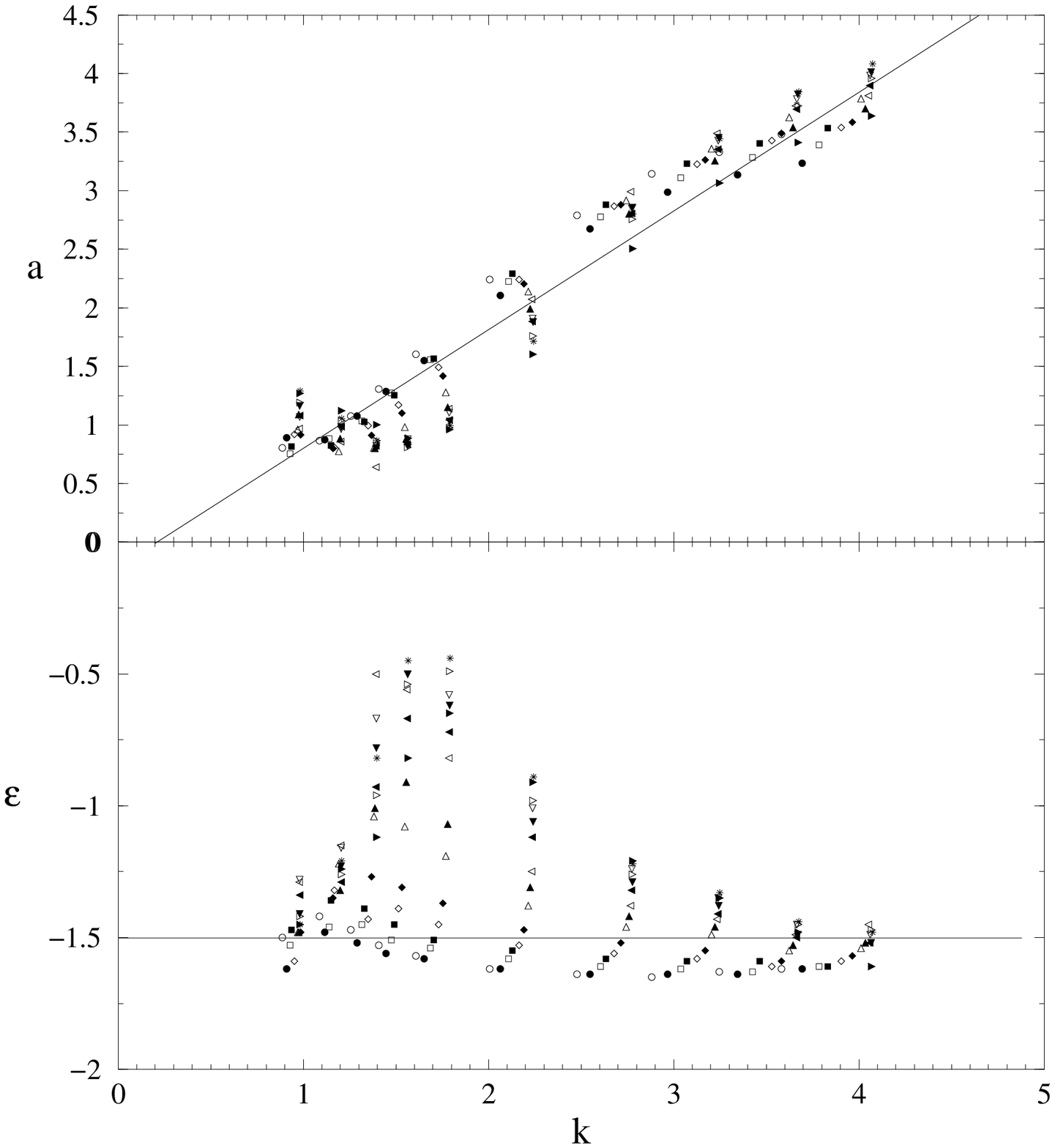}}
\caption{\label{fig-a+epsi+line}
Trend lines for the  parameters $a$ and $\epsilon$  of the simple functional form for the
proton-nucleus partial reaction cross sections.}
\end{figure}

In Fig.~\ref{fig-l0} and ~\ref{fig-a+epsi+line}, we plot the parameter values found for each 
specific energy. Therein, different symbols now indicate different nuclei. 
Empty circles represent the results for ${}^9$Be, filled circles for 
${}^{12}$C, empty squares  for ${}^{16}$O, filled squares  for ${}^{19}$F, empty diamonds for 
${}^{27}$Al, filled diamonds  for ${}^{40}$Ca, empty up triangles  for ${}^{63}$Cu, filled up 
triangles  for ${}^{90}$Zr, empty left triangles for ${}^{118}$Sn, filled left triangles for 
${}^{140}$Ce, empty down triangles  for ${}^{159}$Tb, filled down triangles  for ${}^{181}$Au, 
empty right triangles  for ${}^{197}$Au, filled right triangles  for ${}^{208}$Pb and the stars 
depict the values  for ${}^{238}$U.  The $l_0$ parameter values are plotted as functions
of $kA^{1/3}$ while  the parameters $a$ and $\epsilon$ are plotted as functions of $k$.
It is clear that the parameters vary smoothly with energy.  Of these, the linearity
of $l_0$ with $kA^{1/3}$ is most evident and we find that
\begin{equation}
l_0(E,A)
\sim kR + 2.3\ ;\ \ \ R \sim 1.32 A^{\frac{1}{3}}\ ,
\label{lo-func}
\end{equation}
is a best guess trend.  Although, the variations of $a$ and $\epsilon$ are not as smooth with
$k$, overall they approximate roughly as linear variations with $k$.  The respective trend lines 
are shown in Fig.~\ref{fig-a+epsi+line}.  From this, we find
%%%%%%%%%%%%%%%%
\begin{equation}
a(E,A)
\sim 1.02 k - 0.25\ ; \ \ \ \epsilon \sim -1.5 \,
\label{fn-values}
\end{equation}
to be a simple mean value prescription that might  be used in the integration formulas.
%%%%%%%%%%%%%%%%%%
However the spread of results about these means are significant.  For best results then,
 one should 
consider each parameter set as points in an energy-mass tabulation and  use a  spline or similar 
interpolation to predict results for other energies, and perhaps, other nuclei.  The best fit 
values of these parameters are available in tabular form~\cite{UM02}.  
%%%%%%%%%%%%%%%%%%%%%%%% fig 6 %%%%%%%%%%%%%%%%%%%%%%
%%%%%%%%%%%%%%%%%%%%%%%% fig 7 %%%%%%%%%%%%%%
%%%%%%%%%%%%%%%%%%%

The values of the total reaction cross sections determined using  this simple 
functional form approximation are listed in Tables~\ref{Onetab} and~\ref{Twotab}
as ratios to the actual values that were evaluated with the microscopic optical 
potentials, i.e. the data sums.  
In getting the results shown in the first table ($R_I$), the values of the 
three function parameters found from the fitting program (i.e. the tabulated 
parameter values~\cite{UM02}) were used to give the denominators. 
The results compare well to within a percent in most reflecting the quality 
of fit found by the search process.  The  ratios, $R_{II}$, that are listed in
Table~\ref{Twotab}, were found by using instead the integral results obtained 
using the three parameter sets defined  by the average trend function forms of
Eqs.~\ref{lo-func} and ~\ref{fn-values}.   In those calculations, 
%%%%%%%%%%%%
\begin{equation}
k = \frac{1}{\hbar c} \sqrt{E^2 - m^2 c^4},
\end{equation}
%%%%%%%%%%%%%%%%%%%%%%%%%%%%%%%%%%%%%%%%%%%%%%%
was used in evaluating the denominators with the integral representation for the 
reaction cross sections.  
%%%%%%%%%%%%%%%%%%%%%%%%%%%%%%%%%%%%%%%%%%%%%%%%%%%%%%%%%%%%%%%%%%%%%%%%%%%%%%%%%%%%%%%%%
\begin{table}
\begin{ruledtabular}
\caption{\label{Onetab} Ratios of the total reaction cross sections
formed as sums of the function defined partial cross sections 
and data.}
\begin{tabular}{c|cccccccccc}
Energy (MeV) = & 20 & 30 & 40 & 50 & 65 & 100 & 150 & 200 & 250 & 300\\
\hline
  ${}^9$Be & 1.02 & 1.02 & 1.02 & 1.02 & 1.01 & 1.01 & 1.02 & 1.03 & 1.06 & 1.04 \\
 $^{12}$C  & 1.01 & 1.03 & 1.03 & 1.01 & 1.01 & 1.01 & 1.01 & 1.02 & 1.02 & 1.02 \\
 $^{16}$O  & 1.04 & 1.02 & 1.03 & 1.02 & 1.02 & 1.02 & 1.01 & 1.02 & 1.01 & 1.01 \\
 $^{19}$F  & 1.02 & 1.03 & 1.03 & 1.03 & 1.02 & 1.01 & 1.01 & 1.02 & 1.02 & 1.01 \\
$^{27}$Al  & 1.01 & 1.03 & 1.04 & 1.03 & 1.02 & 1.01 & 1.01 & 1.01 & 1.01 & 1.00 \\
$^{40}$Ca  & 1.01 & 1.03 & 1.03 & 1.03 & 1.02 & 1.02 & 1.01 & 1.01 & 1.00 & 1.00 \\ 
$^{63}$Cu  & 1.01 & 1.02 & 1.02 & 1.03 & 1.03 & 1.02 & 1.02 & 1.01 & 1.01 & 1.00 \\
$^{90}$Zr  & 1.00 & 1.01 & 1.01 & 1.02 & 1.02 & 1.02 & 1.01 & 1.01 & 1.00 & 1.00 \\
$^{118}$Sn & 1.00 & 1.00 & 1.01 & 1.01 & 1.02 & 1.01 & 1.01 & 1.00 & 1.00 & 1.00 \\
$^{140}$Ce & 1.00 & 1.00 & 1.01 & 1.00 & 1.01 & 1.01 & 1.01 & 1.00 & 1.00 & 1.00 \\
$^{159}$Tb & 1.00 & 1.00 & 1.00 & 1.00 & 1.00 & 1.01 & 1.01 & 1.01 & 1.00 & 1.00 \\ 
$^{181}$Ta & 1.00 & 1.00 & 1.00 & 1.00 & 1.00 & 1.01 & 1.01 & 1.00 & 1.00 & 1.00 \\
$^{197}$Au & 1.00 & 1.00 & 1.00 & 1.00 & 1.00 & 1.01 & 1.01 & 1.00 & 1.00 & 1.00 \\ 
$^{208}$Pb & 1.00 & 1.00 & 1.00 & 1.00 & 1.00 & 1.01 & 1.01 & 1.00 & 1.00 & 0.99 \\ 
$^{238}$U  & 1.00 & 1.00 & 1.00 & 1.00 & 1.00 & 1.00 & 1.01 & 1.00 & 1.00 & 1.00 \\
\end{tabular}
\end{ruledtabular}
\end{table}
%%%%%%%%%%%%%%%%%%%%%%%%%%%%%%%%%%%%%%%%%%%%%%%%%%%%%%%%%%%
\begin{table}
\begin{ruledtabular}
\caption{\label{Twotab} Ratios of the total reaction cross sections 
formed using the (n=1) integral representation
and data.}
\begin{tabular}{c|cccccccccc}
Energy (MeV) = & 20 & 30 & 40 & 50 & 65 & 100 & 150 & 200 & 250 & 300\\
\hline
  ${}^9$Be & 1.06 & 0.95 & 0.93 & 0.94 & 0.96 & 0.94 & 1.00 & 1.02 & 1.05 & 1.01 \\
 $^{12}$C  & 1.24 & 1.10 & 1.08 & 1.09 & 1.10 & 1.15 & 1.11 & 1.11 & 1.12 & 1.10 \\
 $^{16}$O  & 1.16 & 1.04 & 1.02 & 1.02 & 1.04 & 1.00 & 1.05 & 1.07 & 1.07 & 1.05 \\
 $^{19}$F  & 1.09 & 0.98 & 0.95 & 0.95 & 0.98 & 0.96 & 1.01 & 1.02 & 1.03 & 1.04 \\
$^{27}$Al  & 1.14 & 1.01 & 1.04 & 0.98 & 1.00 & 0.97 & 1.02 & 1.03 & 1.07 & 1.03 \\
$^{40}$Ca  & 1.17 & 1.03 & 0.98 & 0.99 & 0.98 & 0.97 & 1.02 & 1.03 & 1.04 & 1.03 \\ 
$^{63}$Cu  & 1.19 & 1.00 & 0.94 & 0.95 & 0.93 & 0.92 & 0.96 & 0.98 & 1.00 & 0.99 \\
$^{90}$Zr  & 1.31 & 1.04 & 0.96 & 0.93 & 0.98 & 0.95 & 1.00 & 1.01 & 1.03 & 1.03 \\
$^{118}$Sn & 1.30 & 1.01 & 0.93 & 0.90 & 0.88 & 0.89 & 0.92 & 0.94 & 0.97 & 0.97 \\
$^{140}$Ce & 1.50 & 1.10 & 0.98 & 0.94 & 0.92 & 0.89 & 0.93 & 0.96 & 0.98 & 0.98 \\
$^{159}$Tb & 1.50 & 1.13 & 1.00 & 0.96 & 0.94 & 0.93 & 0.96 & 0.98 & 1.01 & 1.01 \\ 
$^{181}$Ta & 1.75 & 1.22 & 1.05 & 0.99 & 0.96 & 0.94 & 0.97 & 1.00 & 1.02 & 1.02 \\
$^{197}$Au & 1.64 & 1.22 & 1.05 & 0.97 & 0.94 & 0.91 & 0.93 & 0.95 & 0.97 & 0.97 \\ 
$^{208}$Pb & 2.04 & 1.32 & 1.11 & 1.02 & 0.98 & 0.94 & 0.96 & 0.98 & 1.00 & 1.03 \\ 
$^{238}$U  & 2.00 & 1.26 & 1.06 & 0.98 & 0.93 & 0.93 & 0.93 & 0.95 & 0.97 & 0.98 \\
\end{tabular}
\end{ruledtabular}
\end{table}
%%%%%%%%%%%%%%%%%%%%%%%%%%%%%%%%%%%%%%%%%%%%%%%%%%%%%%%%%%%%%%%%%%%%%%%%%%%%%%%%%%%
These ratios indicate that the integral approximation is good to 5\% (ratios in the range 0.95 to
1.05) in some but not all cases. Notably at low energies (20 and 30 MeV) for all masses,
and for almost all energies for ${}^{12}$C, the reproductions are not  good. 
Predominantly this was caused by the choice of linear forms for $a$ and $\epsilon$. 
If on the other hand, the tabled values of the parameters are used in the $n=1$ 
formula of Eq.~\ref{sig_fn_tot}, the ratios again are one to within a few percent.

\section{Conclusions}
The measured  reaction cross sections for 20 to 300 MeV proton scattering from nuclei 
ranging in mass from ${}^9$Be to ${}^{238}$U are very well predicted by the 
calculations made using the $g$-folding model of the optical potential. Those 
calculated results gave a set of partial reaction cross sections in each case (for 
each angular momentum value of import) that vary smoothly with target mass and energy. 
Those variations are well reproduced by a simple 3 parameter function.  The values of 
the parameters ($l_0$, $a$, $\epsilon$) required to set those quality fits also are 
smooth functions of target mass and energy. We believe that such may be used as an 
interpolation tabulation with that simple functional sum to predict partial cross 
sections, and then by summation, the total reaction cross sections of proton-nucleus 
scattering. The extension of using an integral representation to specify the total 
reaction cross sections can be useful but also may diverge from measured values by 
10\% or more if a general linear (with $k$) behavior is adapted for the parameters 
$a$ and $\epsilon$.

%%%%%%%%%%%%%%%%%%%%%%%%%%%%%%%%%%%%%%%%%%%%%%%%%%%%%%%%%%%%%%%%%%%%%%%%

\begin{acknowledgments}
This research was supported by a research grant from the Australian Research Council.
\end{acknowledgments}

%%%%%%%%%%%%%%%%%%%%%%%%%%%%%%%%%%%%%%%%%%%%%%%%%%%%%%%

\bibliography{FN-form}

\begin{thebibliography}{15}
\expandafter\ifx\csname natexlab\endcsname\relax\def\natexlab#1{#1}\fi
\expandafter\ifx\csname bibnamefont\endcsname\relax
  \def\bibnamefont#1{#1}\fi
\expandafter\ifx\csname bibfnamefont\endcsname\relax
  \def\bibfnamefont#1{#1}\fi
\expandafter\ifx\csname citenamefont\endcsname\relax
  \def\citenamefont#1{#1}\fi
\expandafter\ifx\csname url\endcsname\relax
  \def\url#1{\texttt{#1}}\fi
\expandafter\ifx\csname urlprefix\endcsname\relax\def\urlprefix{URL }\fi
\providecommand{\bibinfo}[2]{#2}
\providecommand{\eprint}[2][]{\url{#2}}

\bibitem[{\citenamefont{DiGiacomo et~al.}(1980)\citenamefont{DiGiacomo,
  DeVries, and Peng}}]{Di80}
\bibinfo{author}{\bibfnamefont{N.~J.} \bibnamefont{DiGiacomo}},
  \bibinfo{author}{\bibfnamefont{R.~M.} \bibnamefont{DeVries}},
  \bibnamefont{and} \bibinfo{author}{\bibfnamefont{J.~C.} \bibnamefont{Peng}},
  \bibinfo{journal}{Phys. Rev. Lett.} \textbf{\bibinfo{volume}{45}},
  \bibinfo{pages}{527} (\bibinfo{year}{1980}).

\bibitem[{\citenamefont{Ramsey et~al.}(1998)\citenamefont{Ramsey, Townsend,
  Tripathi, and Cucinotta}}]{Ra98}
\bibinfo{author}{\bibfnamefont{C.~R.} \bibnamefont{Ramsey}},
  \bibinfo{author}{\bibfnamefont{L.~W.} \bibnamefont{Townsend}},
  \bibinfo{author}{\bibfnamefont{R.~K.} \bibnamefont{Tripathi}},
  \bibnamefont{and} \bibinfo{author}{\bibfnamefont{F.~A.}
  \bibnamefont{Cucinotta}}, \bibinfo{journal}{Phys. Rev. C}
  \textbf{\bibinfo{volume}{57}}, \bibinfo{pages}{982} (\bibinfo{year}{1998}).

\bibitem[{\citenamefont{Geissel et~al.}(1995)\citenamefont{Geissel, Munzenberg,
  and Riisager}}]{Ge95}
\bibinfo{author}{\bibfnamefont{H.}~\bibnamefont{Geissel}},
  \bibinfo{author}{\bibfnamefont{G.}~\bibnamefont{Munzenberg}},
  \bibnamefont{and} \bibinfo{author}{\bibfnamefont{K.}~\bibnamefont{Riisager}},
  \bibinfo{journal}{Annu. Rev. Nucl. Part. Sci.} \textbf{\bibinfo{volume}{45}},
  \bibinfo{pages}{163} (\bibinfo{year}{1995}).

\bibitem[{\citenamefont{Wefel}(1987)}]{We87}
\bibinfo{author}{\bibfnamefont{J.~P.} \bibnamefont{Wefel}}, in
  \emph{\bibinfo{booktitle}{NATO ASI Series C: Mathematical and Physical
  Seiences}}, edited by \bibinfo{editor}{\bibfnamefont{M.~M.}
  \bibnamefont{Shapiro}} \bibnamefont{and}
  \bibinfo{editor}{\bibfnamefont{J.~P.} \bibnamefont{Wefel}}
  (\bibinfo{publisher}{Reidel, Dordrecht}, \bibinfo{address}{Erice, Italy},
  \bibinfo{year}{1987}), vol. \bibinfo{volume}{220}.

\bibitem[{\citenamefont{Wilson et~al.}(1982)}]{Wi82}
\bibinfo{author}{\bibfnamefont{J.~W.} \bibnamefont{Wilson}}
  \bibnamefont{et~al.}, in \emph{\bibinfo{booktitle}{NATO ASI Series C:
  Mathematical and Physical Seiences}}, edited by
  \bibinfo{editor}{\bibfnamefont{M.~M.} \bibnamefont{Shapiro}}
  (\bibinfo{publisher}{Reidel, Dordrecht}, \bibinfo{address}{Erice, Italy},
  \bibinfo{year}{1982}), vol. \bibinfo{volume}{107}.

\bibitem[{\citenamefont{Ikeda}(2002)}]{Ik02}
\bibinfo{author}{\bibfnamefont{Y.}~\bibnamefont{Ikeda}}, \bibinfo{journal}{J.
  Nucl. Sci. and Tech. (Japan)}  (\bibinfo{year}{2002}), \bibinfo{note}{in
  press}.

\bibitem[{\citenamefont{Gudowski et~al.}(2002)\citenamefont{Gudowski, Tucek,
  Ericsson, and Walleniu}}]{Gu02}
\bibinfo{author}{\bibfnamefont{W.}~\bibnamefont{Gudowski}},
  \bibinfo{author}{\bibfnamefont{K.}~\bibnamefont{Tucek}},
  \bibinfo{author}{\bibfnamefont{M.}~\bibnamefont{Ericsson}}, \bibnamefont{and}
  \bibinfo{author}{\bibfnamefont{J.}~\bibnamefont{Walleniu}},
  \bibinfo{journal}{J. Nucl. Sci. and Tech. (Japan)}  (\bibinfo{year}{2002}),
  \bibinfo{note}{in press}.

\bibitem[{\citenamefont{Sheng et~al.}(2002)\citenamefont{Sheng, Yanlin,
  Zhixiang, Hongwei, and Zhanglin}}]{Sh02}
\bibinfo{author}{\bibfnamefont{F.}~\bibnamefont{Sheng}},
  \bibinfo{author}{\bibfnamefont{Y.}~\bibnamefont{Yanlin}},
  \bibinfo{author}{\bibfnamefont{Z.}~\bibnamefont{Zhixiang}},
  \bibinfo{author}{\bibfnamefont{Y.}~\bibnamefont{Hongwei}}, \bibnamefont{and}
  \bibinfo{author}{\bibfnamefont{L.}~\bibnamefont{Zhanglin}},
  \bibinfo{journal}{J. Nucl. Sci. and Tech. (Japan)}  (\bibinfo{year}{2002}),
  \bibinfo{note}{in press}.

\bibitem[{\citenamefont{Chadwick}(2000)}]{Ch00}
\bibinfo{author}{\bibfnamefont{M.~B.} \bibnamefont{Chadwick}},
  \emph{\bibinfo{title}{Nuclear Data for Neutron and Proton Radiotherapy and
  for Radiation Protection}}, vol.~\bibinfo{volume}{63} of
  \emph{\bibinfo{series}{ICRU Reports}} (\bibinfo{publisher}{International
  Commission on Radiation Units and Measurements}, \bibinfo{address}{Maryland},
  \bibinfo{year}{2000}).

\bibitem[{\citenamefont{Deb et~al.}(2001)\citenamefont{Deb, Amos, Karataglidis,
  Chadwick, and Madland}}]{De01}
\bibinfo{author}{\bibfnamefont{P.~K.} \bibnamefont{Deb}},
  \bibinfo{author}{\bibfnamefont{K.}~\bibnamefont{Amos}},
  \bibinfo{author}{\bibfnamefont{S.}~\bibnamefont{Karataglidis}},
  \bibinfo{author}{\bibfnamefont{M.~B.} \bibnamefont{Chadwick}},
  \bibnamefont{and} \bibinfo{author}{\bibfnamefont{D.~G.}
  \bibnamefont{Madland}}, \bibinfo{journal}{Phys. Rev. Lett.}
  \textbf{\bibinfo{volume}{86}}, \bibinfo{pages}{3248} (\bibinfo{year}{2001}).

\bibitem[{\citenamefont{Brown}(2000)}]{Br00}
\bibinfo{author}{\bibfnamefont{B.~A.} \bibnamefont{Brown}},
  \bibinfo{journal}{Phys. Rev. Lett.} \textbf{\bibinfo{volume}{85}},
  \bibinfo{pages}{5296} (\bibinfo{year}{2000}).

\bibitem[{\citenamefont{Amos et~al.}(2000)\citenamefont{Amos, Dortmans, von
  Geramb, Karataglidis, and Raynal}}]{Am00}
\bibinfo{author}{\bibfnamefont{K.}~\bibnamefont{Amos}},
  \bibinfo{author}{\bibfnamefont{P.~J.} \bibnamefont{Dortmans}},
  \bibinfo{author}{\bibfnamefont{H.~V.} \bibnamefont{von Geramb}},
  \bibinfo{author}{\bibfnamefont{S.}~\bibnamefont{Karataglidis}},
  \bibnamefont{and} \bibinfo{author}{\bibfnamefont{J.}~\bibnamefont{Raynal}},
  \bibinfo{journal}{Adv. in Nucl. Phys.} \textbf{\bibinfo{volume}{25}},
  \bibinfo{pages}{275} (\bibinfo{year}{2000}).

\bibitem[{\citenamefont{Raynal}(1998)}]{Ra98a}
\bibinfo{author}{\bibfnamefont{J.}~\bibnamefont{Raynal}},
  \emph{\bibinfo{title}{computer program dwba98}} (\bibinfo{year}{1998}).

\bibitem[{\citenamefont{Karataglidis et~al.}(2002)\citenamefont{Karataglidis,
  Amod, Brown, and Deb}}]{Ka02}
\bibinfo{author}{\bibfnamefont{S.}~\bibnamefont{Karataglidis}},
  \bibinfo{author}{\bibfnamefont{K.}~\bibnamefont{Amod}},
  \bibinfo{author}{\bibfnamefont{B.~A.} \bibnamefont{Brown}}, \bibnamefont{and}
  \bibinfo{author}{\bibfnamefont{P.~K.} \bibnamefont{Deb}},
  \bibinfo{journal}{Phys. Rev. C} \textbf{\bibinfo{volume}{65}},
  \bibinfo{pages}{044306} (\bibinfo{year}{2002}).

\bibitem[{\citenamefont{Deb and Amos}(2002)}]{UM02}
\bibinfo{author}{\bibfnamefont{P.~K.} \bibnamefont{Deb}} \bibnamefont{and}
  \bibinfo{author}{\bibfnamefont{K.}~\bibnamefont{Amos}},
  \emph{\bibinfo{title}{Parameter values for a simple functional form for
  proton-nucleus total reaction cross sections}} (\bibinfo{year}{2002}),
  \bibinfo{note}{university of Melbourne preprint, UM-P-008/02}.

\end{thebibliography}

\end{document}